# Large rectification magnetoresistance in nonmagnetic Al/Ge/Al heterojunctions

Kun Zhang,[1] Huan-huan Li,[1] Peter Grünberg,[2] Qiang Li,[1] Sheng-tao Ye,[1] Yu-feng Tian,*,[1] Shi-shen Yan,*,[1] Zhao-jun Lin,[1] Shi-shou Kang,[1] Yan-xue Chen,[1] Guo-lei Liu,[1] & Liang-mo Mei[1]

[1]*School of Physics, State Key Laboratory of Crystal Materials, Shandong University, Jinan 250100, P. R. China*
[2]*Peter Grünberg Institute, Forschungszentrum Jülich, Wilhelm-Johnen-Straße, Jülich, 52428, Germany*

Abstract,

Magnetoresistance and rectification are two fundamental physical properties of heterojunctions and respectively have wide applications in spintronics devices. Being different from the well known various magnetoresistance effects, here we report a brand new large magnetoresistance that can be regarded as rectification magnetoresistance: the application of a pure small sinusoidal alternating-current to the nonmagnetic Al/Ge Schottky heterojunctions can generate a significant direct-current voltage, and this rectification voltage strongly varies with the external magnetic field. We find that the rectification magnetoresistance in Al/Ge Schottky heterojunctions is as large as 250% at room temperature, which is greatly enhanced as compared with the conventional magnetoresistance of 70%. The findings of rectification magnetoresistance open the way to the new nonmagnetic Ge-based spintronics devices of large rectification magnetoresistance at ambient temperature under the alternating-current due to the simultaneous implementation of the rectification and magnetoresistance in the same devices.

Magnetoresistance and rectification are two fundamental physical properties of heterojunctions and respectively have wide applications in spintronics devices. Several types of magnetoresistance (MR) such as anisotropic MR,[1] giant MR,[2] and tunneling MR[3] are presently indispensable in current data storage technology. Later, the observation of emergent magnetoresistance phenomenon such as spin Hall MR[4,5] and large MR in nonmagnetic materials[6-9] has inspired further investigations aiming to harness the spins of electrons rather than just their charges in the next-generation spintronic devices. On the other hand, rectification of high frequency alternating-current (AC) or detection of high frequency signals by conversion into a direct-current (DC) signal is typically achieved by using Schottky diodes or semiconducting p-n junctions. Beside this conventional functionality, new concept of rectifications[10-15] have been developed such as spin rectification in magnetic tunnel junctions[10,11], giant MR stripes[12], and anisotropic MR microstrips[13-15] to pave the way for designing new spin sources for spintronic applications. Up to date, the study of magnetoresistance and rectification effects are relatively independent to each other. This leaves our understanding of the interplay between electrostatic response and spin dynamics incomplete and limits the further development of spintronics.

In this work, we report a very different magnetoresistance in nonmagnetic Al/Ge Schottky heterojunctions that can be regarded as *rectification magnetoresistance* (RMR), which is due to the simultaneous implementation of the rectification and magnetoresistance in the same devices. It is found that the application of a pure small sinusoidal alternating-current to the nonmagnetic Al/Ge Schottky heterojunctions can generate a significant DC voltage, and this rectification voltage strongly varies with the external magnetic field. A rectification magnetoresistance as large as 250% is observed in Al/Ge Schottky heterojunctions at room temperature while the conventional magnetoresistance is only of 70% in the same devices. The findings of rectification magnetoresistance open an alternative way towards novel nonmagnetic Ge-based spintronics devices.

**Results**

**Observation of rectification magnetoresistance.** The electrical transport experiments were performed on Al/Ge/Al heterojunctions in a circle configuration as schematically shown in the inset of Fig. 1a, where the center Al/Ge electrode is designed to be Schottky contact while the circle Al/Ge electrode is Ohmic contact at room temperature. Here the Ge substrate is intrinsic semiconductor. In Fig. 1a, non-linear I-V curves has been observed at 300 K, which is a signature that band bending of the junction interface leads to the formation of the center Al/Ge Schottky barrier. A careful investigation shows that the I-V curve is asymmetric about zero voltage, further indicating the existence of rectifying behavior of Al/Ge/Al heterojunctions. Moreover, when the external magnetic field increases from 0 to 6

Tesla, the voltage at a fixed current monotonically increases as shown in Fig. 1a, resulting in a positive magnetoresistance of 70% under the magnetic field of 6T at the direct-current of 10 μA as shown in Fig. 1b.

In order to explore the effects of the simultaneous implementation of the rectification and magnetoresistance in the same devices, we applied a pure sinusoidal alternating-current to the devices and measured the magnetic field dependence of the rectifying DC voltage, i.e., rectification magnetoresistance. Here, the MR is always defined as MR=$(V_H-V_0)/V_0 \times 100\%$, where $V_H$ and $V_0$ are respectively the detected DC voltage with and without magnetic field at a fixed DC for measuring conventional MR or a pure sinusoidal AC for measuring rectification MR. The most intriguing and innovative part of present investigation is the observation of greatly enhanced rectification MR as shown in Fig. 1c. We find that the rectification MR in Al/Ge/Al heterojunctions is as large as 250% at room temperature for the AC with the amplitude of 50 μA and frequency 1000 Hz, which is greatly enhanced as compared with the conventional MR of 70%.

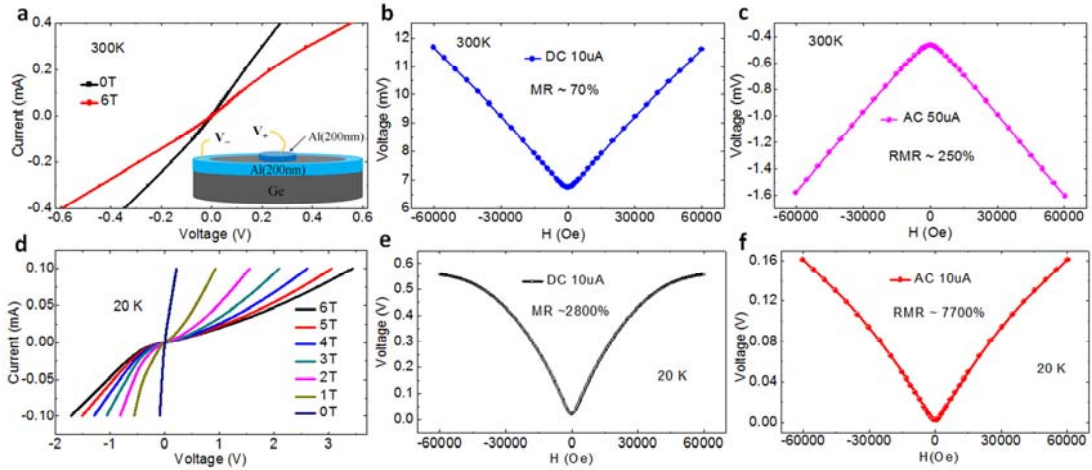

Figure 1. **Rectification magnetoresistance and conventional magnetoresistance of the Al/Ge/Al heterojunctions. a-c**, respectively, shows the I-V curves, conventional MR, and rectification MR of the Al/Ge/Al heterojunctions in a circle configuration as schematically shown in the inset of Fig. 1a, which were measured at 300 K. **d-f**, respectively, shows the I-V curves, conventional MR, and rectification MR, which were measured at 20 K.

The low temperature electrical transport properties were further measured on Al/Ge/Al heterojunctions. At low temperature such as 20 K, both the center Al/Ge and circle Al/Ge electrodes become Schottky contacts. In this case, I-V curves become more non-linear and asymmetric (Fig. 1d), leading to larger rectification MR of 7700% under 10 μA applied AC (Fig. 1f), while the conventional MR is only 2800% (Fig. 1e). Moreover, due to the difference in asymmetry of the I-V curves at different temperatures, the rectifying DC voltage and the conventional DC voltage have the same sign at the low temperature 20 K, while they have opposite sign at 300 K.

**Current amplitude and frequency dependence of rectification magnetoresistance.** Fig. 2a shows the current amplitude dependence of the rectification MR of the Al/Ge/Al heterojunctions in the circle configuration measured at 20 K for the fixed frequency 1000 Hz. It is clear that the rectification MR is much larger than the conventional MR in the current range of 5 μA to 100 μA. Fig. 2b shows the frequency dependence of the rectification MR measured at 300 K for a similar Al/Ge/Al heterojunctions. It reveals that the rectification MR shows a very weak frequency dependence in the low frequency range (<2 KHz), while it decreases quickly with increasing frequency.

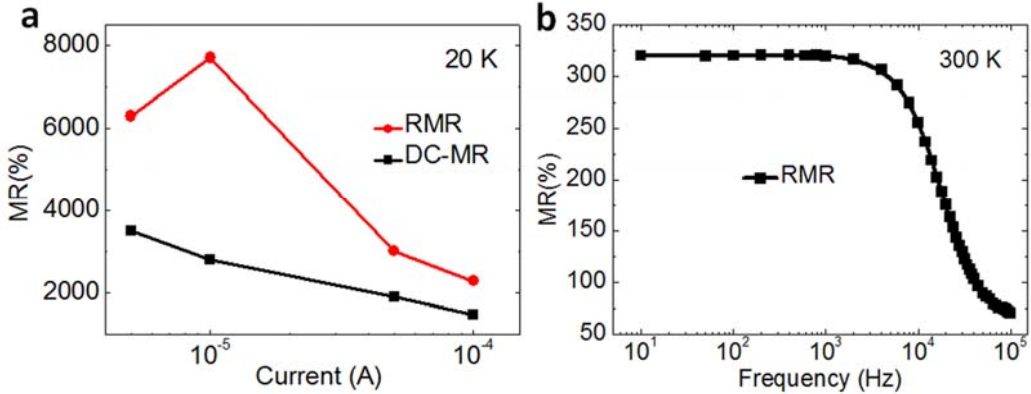

Figure 2. **Current amplitude and frequency dependence of the rectification magnetoresistance. a.** The current amplitude dependence of the rectification MR (marked as RMR) of the Al/Ge/Al heterojunctions in the circle configuration measured at 20 K for the fixed frequency 1000 Hz. As comparison, the conventional MR (marked as DC-MR) was shown, which was measured by using DC with the same current amplitude as AC. **b.** The frequency dependence of the rectification MR for another similar Al/Ge/Al heterojunction in the circle configuration measured at 300 K under AC current of 10 μA.

**Control experiments of rectification magnetoresistance.** To confirm the rectification MR, the following reference experiments have been performed. First, we prepared Al/Ge/Al heterojunctions in a bar configuration as schematically shown in the inset of Fig. 3a, where both the Al/Ge bar electrodes are designed to be Schottky contacts in series but with opposite rectification direction. The non-linear asymmetric I-V curves, 80% conventional MR, and 200% rectification MR are found in this configuration of two Schottky heterojunctions at room temperature, as shown in Figs. 3a-c. Second, we prepared the In/Ge/In Schottky contacts on the n-type Ge (single side polished, <111> orientation with resistivity of 0.02 - 0.15 Ωcm), which have obvious rectification effect (Fig. 3d) but negligible conventional MR (Fig. 3e). These In/Ge/In Schottky contacts indicate that only rectification effect without conventional magnetoresistance cannot induce any rectification MR, as shown in Fig. 3f. Third, we prepared the $Co_{0.7}Zn_{0.3}O$ magnetic semiconductor films with obvious conventional MR (Fig. 3e) but negligible rectification effect (Fig. 3d). This negative MR is a result of spin dependent hopping[16]. It is obvious that only conventional magnetoresistance without rectification effect can not induce any rectification MR either, as shown in Fig. 3f. Fourth, we simply combine the In/Ge/In Schottky contacts and the $Co_{0.7}Zn_{0.3}O$ films in series, which have obvious rectification effect (Fig. 3d) and conventional MR (Fig. 3e). However, the above combined device can not induce any rectification MR, as shown in Fig. 3f. This unambiguously indicates that a simple combination of rectification device and magnetoresistance device can not induce any rectification MR. Furthermore, it is worthy to mention that any pure resistance devices with/without magnetoresistance in series with rectification MR devices do not produce any additional rectifying DC voltage, which is very beneficial to the application of the rectification MR devices.

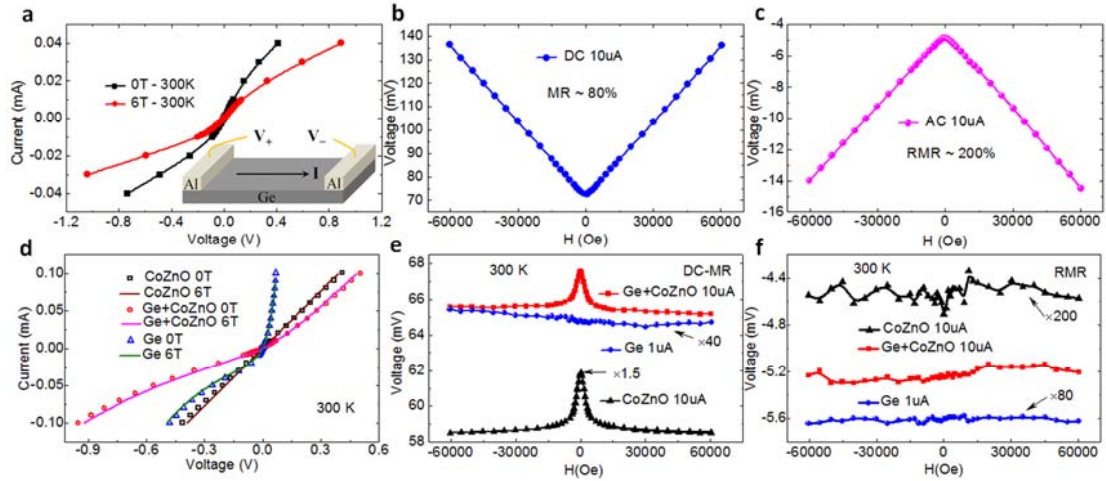

Figure 3. **Rectification magnetoresistance and conventional magnetoresistance of four reference samples. a-c**, respectively, shows the I-V curves, conventional MR, and rectification MR of the Al/Ge/Al heterojunctions in a bar configuration as schematically shown in the inset of Fig. 3a. **d-f**, respectively, indicates the I-V curves, conventional MR, and rectification MR of the reference samples In/Ge/In Schottky contacts (marked as Ge), $Co_{0.7}Zn_{0.3}O$ magnetic semiconductor films (marked as CoZnO), and the combination of the In/Ge/In Schottky contacts and the $Co_{0.7}Zn_{0.3}O$ films in series (marked as Ge+CoZnO).

**Mechanisms of rectification magnetoresistance.** Now we discuss the possible mechanisms of the conventional MR and rectification MR in the Al/Ge/Al heterojunctions. A few mechanisms are known to induce positive magnetoresistance in semiconductor diodes and nonmagnetic materials. The large positive MR is a characteristic of the inhomogeneous semiconductors and p-n junctions[6,8,17-19]. For instance, a large positive MR could be induced as a result of electric-field inhomogeneity caused by space-charge effect[8], where quasi-neutrality breaking need to be achieved under large applied electric voltage as high as several tens volts. In our case, however, the positive MR is observed at several millivolts. The measuring voltage of several millivolts is three or four orders lower than reported voltage needed to show obvious space charge effect. Moreover, the I-V curve of the studied Schottky diodes is near linear in this low voltage range, which is different from the I-V curves of the space charge effect. Therefore, the space-charge effect can not explain the conventional MR in our Al/Ge/Al heterojunctions. On the other hand, large positive MR was explained by spin-dependent electron filling in the interface band structures in magnetic semiconductor diodes or magnetic p-n junctions[20-22], and it was explained by diode-assisted MR mechanism in diode-assisted Ge, Si and GaAs nonmagnetic semiconductors[23,24]. Obviously, these mechanisms can not explain the conventional MR in our Al/Ge/Al heterojunctions either.

In line with above-mentioned effect, the shrinking of the wave function of the impurity state could lead to a large non-saturating positive magnetoresistance in lightly doped semiconductor materials and artificial structures as well[7, 25-28]. Considering the fact that intrinsic Ge devices with a lower carrier concentration showed a much significant MR compared to the n-type Ge devices, we believe that quantization of the carrier motion by the magnetic field is responsible for the present positive MR[25]. In particular, for small enough carrier concentration, only a finite overlap exists between the electron wave functions of impurity and/or interfacial states. The application of magnetic field causes the shrinkage of the carrier wave function and the overlap of wave-function "tails" between different states is greatly reduced. In such a way, magnetic field gradually narrows the band width of impurity and/or interfacial states and increases it to a slightly higher energy[7,25]. This slight increase of band energy due to the magnetic field further modifies the energy band bending at the Al/Ge Schottky interfaces, which leads to a large positive MR.

The mechanism of the rectification MR is related to both conventional MR and rectification effect of Schottky heterojunctions. As previously mentioned, both conventional MR and rectification MR are defined as MR=$(V_H-V_0)/V_0 \times 100\%$. By introducing a pure sinusoidal AC, $I = I_0 \sin(\omega t)$, into the Schottky heterojunctions to measure the rectification MR, the detected DC voltage $V_H$ at any specific magnetic field equals to the average of the corresponding real time voltage $V_H(t)$, i.e., $V_H = \frac{1}{T}\int_0^T V_H(t)dt$. Here $T$ is the time period of the applied AC current. For the same reason, we can obtain $V_0 = \frac{1}{T}\int_0^T V_0(t)dt$ and further the rectification MR was obtained. It is reasonable to believe that $V_H(t)$ has the same magnetic field dependence as conventional MR for relatively low AC frequency, i.e. the wave function shrinking mechanism. Despite this similarity, the rectification MR is essentially different from the conventional MR. On one hand, the rectification MR can be much larger than the conventional MR; on the other hand, without rectification effect, the rectification MR disappears no matter whether the device has conventional MR or not. Up to date, though the most relevant physics behind rectification MR effect has been captured, further quantitative and extended investigations are required to clarifying the current amplitude and frequency dependence of the rectification MR.

**Discussion**

The present rectification MR should be observable in other Schottky diodes and/or p-n junctions with both rectification and magnetoresistance simultaneously. One can readily anticipate devices with even higher rectification MR values than reported in this proof of principle investigation by optimizing the device design. Nevertheless, this amazing rectification MR accompanied with high sensitivity definitely provides us an alternative way towards advanced germanium-based magnetoelectronics technologies.

**Methods**

**Sample preparation.** The experiments were performed on Al/Ge/Al heterojunctions in a circle configuration as schematically shown in the inset of Fig. 1a, where the center Al/Ge electrode is designed to be Schottky contact while the circle Al/Ge electrode is Ohmic contact at room temperature. The Ge substrate is intrinsic semiconductor (single side polished, <100> orientation with resistivity of 55.6 - 59.4 Ωcm). The circle Al/Ge electrode was annealed at 550°C for 3 minutes to form Ohmic contact at room temperature. Then the center Al/Ge electrode with the diameter of 150 um is designed to be Schottky contact. The separation distance between the two electrodes is 150 um. The Al/Ge/Al heterojunctions in a bar configuration are 1 mm in length and 150 um in width. Both Al/Ge electrodes are connected by Al wire bonding at different temperature and pressure to obtain two different Schottky contacts in series but with opposite rectification direction.

**Transport measurements.** The I-V curves and conventional MR were measured with Keithley 2400 electrical current source meter and Keithley 2182 voltage meter. The rectification MR was measured with Keithley 6221 source meter to provide a sinusoidal AC input and Keithley 2182 voltage meter to measure the generated DC voltage. For all the MR measurements described above, the external magnetic field was applied in the film plane.

**Acknowledgements**

We acknowledge the financial support from the NBRP of China no. 2013CB922303 and 2015CB921502, the key program of NSFC no. 11434006, the NSFC for Distinguished Young Scholar no. 51125004, 111 project no. B13029, and NSFC no. 11374187.

**Author information** Reprints and permissions information is available at npg.nature.com/reprintsandpermissions. The authors declare no competing financial interests. Correspondence and requests for materials should be addressed to Y.F.T. and S.S.Y. (yftian@sdu.edu.cn; shishenyan@sdu.edu.cn).